# An ion species model for positive ion sources – part II analysis of hydrogen isotope effects

E Surrey[1] and A J T Holmes[2]
[1]EURATOM/CCFE Association, Culham Science Centre, Abingdon, Oxfordshire OX14 3DB, UK
[2]Marcham Scientific, Hungerford, Berkshire RG17 0LH, UK

E-mail:elizabeth.surrey@ccfe.ac.uk

**Abstract.** A one dimensional model of the magnetic multipole volume plasma source has been developed for application to intense ion/neutral atom beam injectors. The model uses plasma transport coefficients for particle and energy flow to create a detailed description of the plasma parameters along an axis parallel to that of the extracted beam. In this paper the isotopic modelling of positive hydrogenic ions is considered and compared with experimental data from the neutral beam injectors of the Joint European Torus. The use of the code to gain insights into the processes contributing to the ratios of the ionic species is demonstrated and the conclusion is drawn that 75% of the atomic ion species arises from ionization of dissociated molecules and 25% from dissociation of the molecular ions. However whilst the former process is independent of the filter field, the latter is sensitive to the change in distribution of fast and thermal electrons produced by the magnetic filter field and an optimum combination of field strength and depth exists. Finally the code is used to predict the species ratios for the JET source operating in tritium and hence the neutral beam power injected into the plasma in the JET tritium campaign planned for 2016.

## 1. Introduction

Neutral beam injection is a major method of providing non-inductive current drive and heating to magnetically confined fusion plasma devices. The beams, generally of hydrogenic atoms, are formed from either positive or negative ion precursors, the choice being dictated by the beam energy required for plasma penetration. For positive ions, the neutral beam is created by charge exchange in a gas cell for which the neutralisation efficiency is less than 30% at beam energy above ~80keV/amu. For negative ions the neutral beam is created by electron stripping in a gas cell, for which the efficiency has a maximum value of ~58% for all energies above ~50keV/amu. The plasma generator that forms the precursor ions is of the magnetic multipole type and may contain a magnetic "filter" to create a low temperature plasma in the region from which the beam is extracted. This technique, which reduces the electron energy in the extraction region, has different consequences for positive and negative ion production. For positive ion sources it serves to reduce the direct ionisation of the neutral gas, limiting the $H_2^+$ production and promoting the dissociation of $H_3^+$ into $H^+$ and $H_2$ [1]. (Here the use of H will be taken to represent all hydrogen isotopes unless specifically stated). In the negative ion source the electron energy is reduced to an energy at which electron detachment collisions are no longer dominant [2] so that the production of $H^-$ is maximised. Both of these effects occur at an electron temperature of approximately 1eV. The negative ions are formed by either dissociative attachment collisions (also maximised at about 1eV temperature) or by emission from a caesiated surface.

Despite its history the role of this magnetic filter is not clearly understood and a model developed for this purpose was described in the first paper of this series [3]. The magnetic filter itself is extremely simple; it is a sheet of magnetic field dividing the source chamber into two essentially field free regions - the one next to the beam extraction apertures referred to as the extraction region and the other at the back of the source chamber referred to as the driver region. This latter region also contains the discharge excitation system. This can be an RF drive antennae but for many discharges it is a DC hot wire filament. This paper will focus on the latter technique as most of the data presented here is from sources of this type.

The first use of a magnetic filter was Ehlers and Leung [4] and Holmes et al [5,6,7] where it was used to enhance the production of $H^+$ ions relative to the molecular ions $H_2^+$ and $H_3^+$. The molecular ions are destroyed by cold electron impact while the lack of fast electrons, blocked by the magnetic filter, prevents ionization and the reforming of new molecular ions. Successful magnetic filters were developed for the PINI ion source of the neutral beam injectors of the Joint European Torus (JET) [6,7] and also for the "10×10" source developed at Lawrence Berkeley Laboratory by Pincosy et al [8]. However despite considerable experimentation, a model that could explain in detail how these sources operate was never achieved although a general understanding of the filter was achieved [6]. The only significant difference between the sources used for proton (or deuteron) enhancement and those where $H^-$ or $D^-$ production is required is the reversal of the accelerator polarity and design and the use of a slightly stronger magnetic filter [9] where the field is roughly doubled. Holmes[10] developed a model of the magnetic filter in its role in $H^-$ production, but no reference was made to positive ion species.

This paper expands upon the application of the model developed in Part I [3] to include examination of the effect of hydrogenic isotope and the magnetic filter field. The results of modelling the species ratios for hydrogen and deuterium plasmas are compared with experimental data and the recombination and energy accommodation coefficients adjusted for optimum agreement. The model is then used to explore the influence of the magnetic filter field on the species ratio and source efficiency, revealing a subtle dependency on the plasma distribution created by the filter. Finally the model is used to predict the species ratios form a source operated in tritium for the future D-T campaign at the Joint European Torus, planned for 2016-17.

**2. The Model Concept**
The model was described in detail in Part I [3] and will only be summarised here. It is a major extension of the earlier one dimensional model of a negative ion source by Holmes [10], modified to represent the primaries as a one dimensional fluid in the same form as the plasma electrons and ions in order to include the species evolution and the calculation of atomic density through the source. This allows the description of poorly confined sources such as the pancake source described by Ehlers and Leung [11], which forms an almost pure $H_2^+$ discharge where there is a large loss of primaries that are mobile throughout the entire source volume. An earlier version of the model described here was successfully used to analyse the data from that experiment with the intention of creating a source of $H_2^+$ ions that could simulate the effects of $D^+$ acceleration in the RFQ and subsequent drift tube linac for the IFMIF project [12].

The model is illustrated in simplified flow chart form in figure 1. Before the plasma transport is examined, the primary flux and energy flow needed to sustain the plasma is established. The plasma modelling is divided into three parts, the primary electron input, the plasma itself and finally the system at the plasma grid, which normally floats at a negative potential with respect to the plasma and source wall (anode) due to the influx of primaries. In most sources there is zero current to this grid as the ion and total electron currents cancel. The code however retains the possibility of applying an external potential bias and positive currents are defined as net ion flow to the grid.

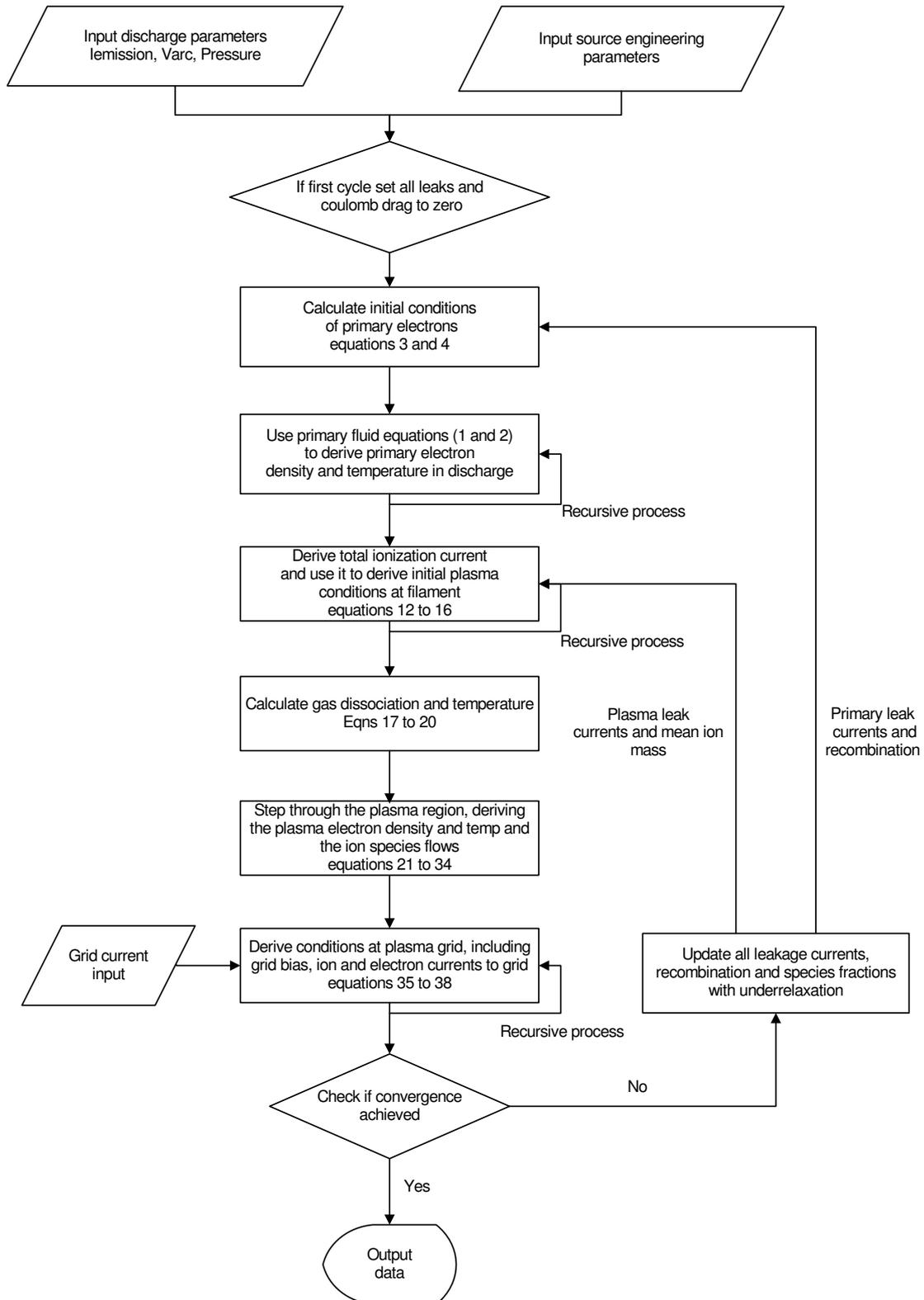

**Figure 1** Flow diagram of the model, simplified for clarity.

The primary electrons emitted from the filament transfer their energy through inelastic collisions and escape the system at the anode (which is magnetically confined) or the plasma grid. The plasma electrons, degraded primaries and ions are also lost on the anode and plasma grid but there is an additional constraint at the latter whereby the algebraic sum of the electron and ion currents is equal to the grid current. The ion flux in turn is defined by the ionization rate less any losses to anode or filament and matches the ion flux surviving the transit across the plasma. However the ion losses to the walls and filaments are not known at the start of the calculation and are found by iteration.

The basic structure is to initially create a very simple model of the plasma without plasma losses, coulomb drag or $H_3^+$ ions. This allows the formation of a preliminary primary electron distribution which is used to create a plasma distribution. Then first estimates of the losses and coulomb drag are made together with the revised ion species distribution. This information, suitably relaxed to avoid numerical oscillation, is fed back into the primary distribution and the whole process is repeated until convergence is achieved. With this approach the model can incorporate almost all the physics known to exist inside a bucket source with as much detail as desired. All the collision processes, listed in table 1, are temperature dependent and this is included via suitable fitting functions, so that the local rate process depends on the local electron temperature. For convenience the data input is divided into discharge variables (i.e. arc voltage, current and pressure) and "engineering" variables which describe the source chamber (i.e. cusp field, filament position, plasma grid area etc.)

**Table 1.** Analytic forms of reaction rates used in the model.

| Process | label | Expression in $m^3.sec^{-1}$ |
|---|---|---|
| $H_2 + e \rightarrow 2H + e$ | xs2 | $xs2 = 10^{-15}\exp(3.2)\exp(-8/T)F$ <br> if T>22, then F = 1 <br> else F = exp(-0.068(T-22)) |
| $H_2^+ + e \rightarrow 2H$ | xs4 | $xs4 = 10^{-15}\exp(3.5)(1-\exp(-5.2/T))$ |
| $H_2^+ + e \rightarrow H^+ + H + e$ | xs5 | $xs5 = 10^{-15}\exp(4.5)\exp(-2.5/T)$ |
| $H_2^+ + H_2 \rightarrow H_3^+ + H$ | xs6 | $2.1 \times 10^{-15}$ |
| $H_3^+ + e \rightarrow H^+ + H_2 + e$ | xs7 | $xs7 = 10^{-15}\exp(6.5)\exp(-13/T)$ |
| $H_3^+ + e \rightarrow H_2 + H$ | xs8 | $xs8 = 10^{-15}\exp(3.8)(1-\exp(-6/T))$ |
| $H_2 + e \rightarrow H_2^+ + 2e$ | $q_i$ | $q_i = 10^{-15}\exp(3.98)\exp(-22.84/T)$ |

The ions are heated by energy transfer from the electrons by coulomb collisions and these create hot atoms and molecules on colliding with the wall. These atoms and molecules elevate the global gas temperature and the efficiency of this process is described by two recombination coefficients, $\gamma$, the atomic hydrogen recombination rate and $\gamma_h$, the thermal accommodation rate. The values are not known with any certitude and can only be determined by comparing the code results with experimental results. However Chan et al [13] argue that $\gamma$ should be about 0.2 for deuterium discharges although experimental data from Wood and Wise [14] suggest it should a little smaller. The model can be used to undertake a sensitivity analysis to these coefficients and this will be discussed in Section 3.3.

**3. Isotope Scaling Effects – Comparison of the Model with Experimental Results**

In Part I [3] the model was compared to experimental results from two plasma generators: the "10×10" source developed by Pincosy et al [8] at the Lawrence Berkeley Laboratory for neutral beam heating and the "PINI" source used for neutral beam heating [15] on the JET experiment, for which extensive ion species measurements with and without a filter field exist. The comparison was limited to operation in deuterium for which extensive databases exist for both sources. For the JET PINI there is also an extensive positive ion species database in hydrogen, including with a filter field strong enough to allow it to act as a negative ion source, hence bridging the gap between the two modes of operation. This section therefore considers only the JET PINI source, comparing data from operation in both hydrogen and deuterium and the results of the model. The details of magnetic filter field configuration are given in paper 1 but the results presented here relate to the JET PINI with a super cusp filter field and with no filter field.

Modelling the isotope effects requires knowledge of those processes that have a mass dependency. The majority of the dissociation and ionization processes in the model, shown in table 1, are electronic and it is assumed that there is no mass dependency. The only exception is the reaction $H_2^+ + H_2 \rightarrow H_3^+ + H$ for which the rate coefficient is constant in the model. The only remaining isotopic dependence is the two wall recombination coefficients, $\gamma$, the atomic hydrogen recombination rate and $\gamma_h$, the thermal accommodation rate.

The values are not known with any certitude as they depend upon the surface condition of the wall (which will be a mixture of adhered gas on metal, mainly tungsten from the filaments but possible with some copper). Other effects such as temperature and roughness will also influence these values. The effects of absorbed gas and surface temperature were accommodated in an analysis by Song [16] based on experimental data which gives a general expression for any gas-surface combination. Using this analysis the accommodation coefficient $\gamma_h$ was calculated for atomic and molecular forms of the three isotopes as shown in table 2 for both atomic and molecular neutral species for a tungsten wall at a temperature of 350K

Table 2. Values of Accommodation Coefficient calculated from the analysis of Song [16].

| Monatomic | $\gamma_h$ | Diatomic | $\gamma_h$ |
|---|---|---|---|
| H | 0.11 | $H_2$ | 0.25 |
| D | 0.20 | $D_2$ | 0.39 |
| T | 0.27 | $T_2$ | 0.49 |

The values of $\gamma$ and $\gamma_h$ for hydrogen and deuterium were finally determined by best fit to the data but were not modified for the different filter configuration simulations and are given in table 3. No attempt has been made in the model to differentiate between monatomic and diatomic thermal accommodation, so the values of $\gamma_h$ would be expected to fall between the two limits in table 2, as indeed is the case. The values of $\gamma$ for deuterium are commensurate with those in [14]

Table 3. Values of Accommodation Coefficient derived from best fit to the data.

| | $\gamma$ | $\gamma_h$ |
|---|---|---|
| H | 0.05 | 0.2 |
| D | 0.13 | 0.35 |

*3.1 JET PINI with no filter field*

The data and model results for hydrogen and deuterium are shown in figure 2 for the filterless source. The most obvious feature is the lower yield of $H^+$ compared to $D^+$ at the same current density at the extraction plane. The standard explanation has been that this is due to the difference in drift velocity of the two isotopes, but the ratio of the full energy species ($H^+:D^+$) would then be equal to the ratio of the square root of the molecular masses i.e. $\sqrt{2}=0.707$. In fact the $H^+:D^+$ ratio is larger than this and is almost constant over the range of current density shown with an average value of 0.90. As there is no filter field to modify the electron transport and energy and the electron temperature is similar for both gases, this difference must reflect different values of $\gamma$ and $\gamma_h$ required in the model to match the experimental data and their effects on the atomic production.

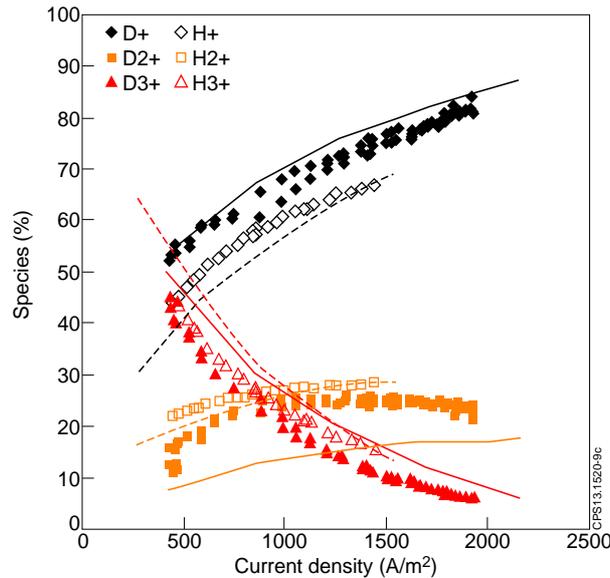

**Figure 2** Comparison of species ratios in hydrogen and deuterium for the filterless source. The open symbols are experimental data in hydrogen, the closed symbols in deuterium as shown in the key. The broken curves are modelling for hydrogen, the solid curves for deuterium.

*3.2    The JET PINI source with supercusp filter*
The equivalent isotope comparison for the supercusp source is shown in figure 3. Again the ratio of the proton to deuteron fraction is higher than the simple velocity scaling and is almost constant with current density, having an average value of 0.86, slightly lower than for the filterless source but is probably not significant. Again the accommodation coefficients are probably the cause for this.

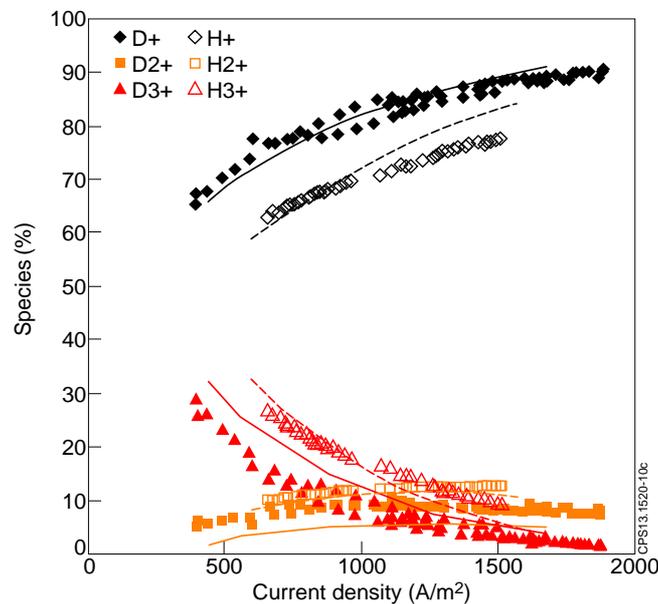

**Figure 3** Comparison of species ratios in hydrogen and deuterium for the supercusp filter source. The open symbols are experimental data in hydrogen, the closed symbols in deuterium as shown in the key. The broken curves are modelling for hydrogen, the solid curves for deuterium.

*3.3 Consideration of the accommodation coefficients*
The accommodation coefficients, $\gamma$ and $\gamma_h$, have been shown to be influential in determining the species ratios in the two types of source and given the uncertainty in their values, a sensitivity study for $\gamma$ and $\gamma_h$ is shown in figure 4 for the filterless source. The species ratios are more sensitive to changes in $\gamma$ but show opposite trends for the two different coefficients.

To first order, it would seem that small errors in $\gamma$ and $\gamma_h$ are unlikely to yield large inconsistencies in the species ratio when extended to tritium.

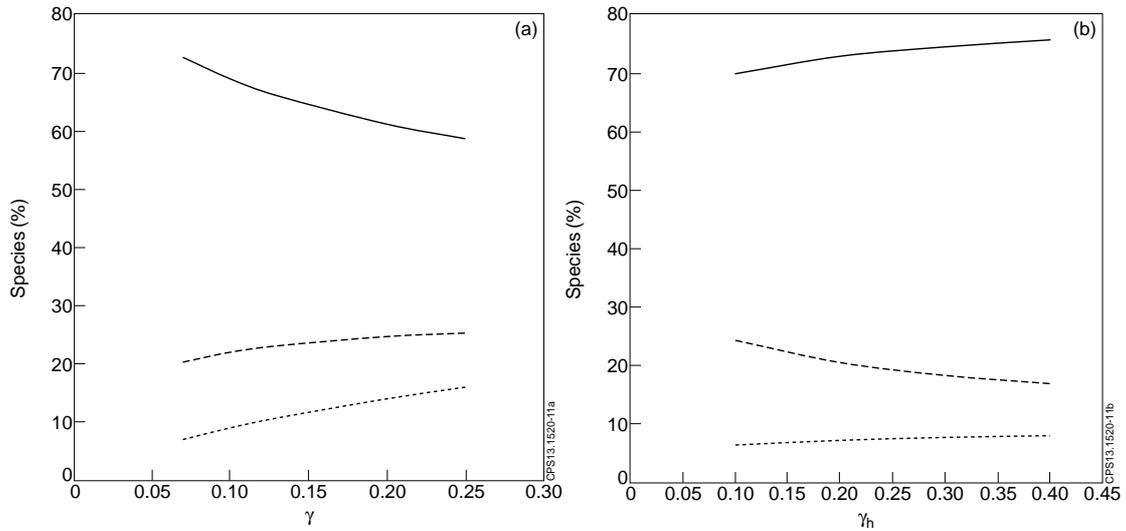

**Figure 4** Sensitivity of species ratio to (a) $\gamma$ and (b) $\gamma_h$ in hydrogen. Arc current 1000A, source pressure 0.36Pa. $H^+$ —, $H_2^+$ – – –, $H_3^+$ - - -

### 3.4 *Modelling of the isotope dependency of the species ratio*

It is shown that the model reproduces the general trends of the isotope dependency of the species ratio in both filterless and supercusp sources, providing the accommodation coefficients are regarded as parameters of the isotope mass. The agreement with the experimental values is variable from 1% error to 70% in the worst case. (Interestingly the worst agreement is always found for the diatomic molecular ion and this data is generally regarded as having the greatest uncertainty due to the analysis technique). The error in the measurements for $H^+$ and $H_3^+$ is approximately 10% due to cross section uncertainty and the assumption of a known neutralization target and approximately 20% for $H_2^+$.

## 4 Mechanism of proton enhancement by filter field

In earlier work on the reasons for the enhancement by a filter of the proton (deuteron) yield, it was asserted that the main cause was the break-up of the molecular ions and the prevention of formation of new molecular ions in the extraction region where the electron energy is too low to be effective. A filter would enhance this, and does in this model, but the whole process is rendered more complicated by the impact the filter has on the overall source efficiency. This is seen in figure 5 where current density at the extraction plane is plotted as a function of filter field strength for a constant arc current.

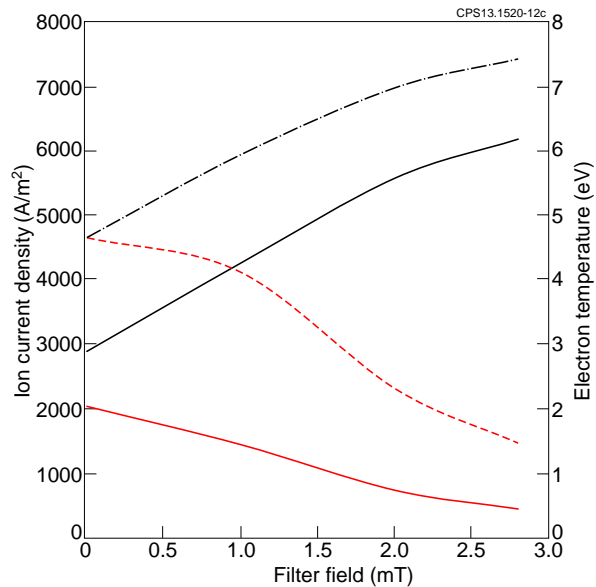

**Figure 5** The current density and electron temperature in the driver and extraction regions as functions of filter field strength. Driver region ━ current density - - - electron temperature; extraction region ━ current density - - - electron temperature.

Initially the driver and extraction regions are very similar but soon show a significant difference at fields in excess of one millitesla. The reduction in plasma current density in the extraction region is due to the reduced transport of the plasma across the filter field. In the driver region, the plasma density tends towards saturation, as shown in figure 5, with increasing filter field due to losses of plasma along the magnetic confinement cusp lines. Similar arguments pertain to the electron temperature in both parts of the source: the filter field reduces energy transport into the extraction region, leading to a reduced electron temperature, whilst the increased cusp losses result in increased energy loss from the driver region. The high plasma density in the driver region has other effects apart from increased cusp losses: the gas is partially converted to the atomic form, allowing direct formation of protons, and molecular ion break-up, which scales as the square of the plasma density, is much enhanced. The previous theory for this effect simply argued that the transit of the molecular ions across the extraction region was sufficient to destroy most of them and the lack of primary ionization here prevented new molecular ions from forming.

In general, part of the proton (deuteron) production arises from direct ionization of the atomic gas, part from dissociation of the molecular ions and finally part from the redistribution of the plasma density and temperature when a filter is added. The contributions of these different effects can be investigated by using the model to isolate atomic and molecular contributions as discussed in the following sections.

*4.1 Molecular Dissociation in the Extraction region*
Previous attempts to increase the proton yield, based on the molecular break-up hypothesis, involved increasing the length of the source between the extraction and filter field planes (Pincosy et al [8]). From the arguments above the idea of making the source deeper has little merit except if there is a need to have a high proton fraction at low current density. In figure 6 the proton fractions of a PINI source with 0.2, 0.27 and 0.34 meters depth are compared, the filter staying at the same position relative to the source backplane so that it is the extraction region that is enlarged.

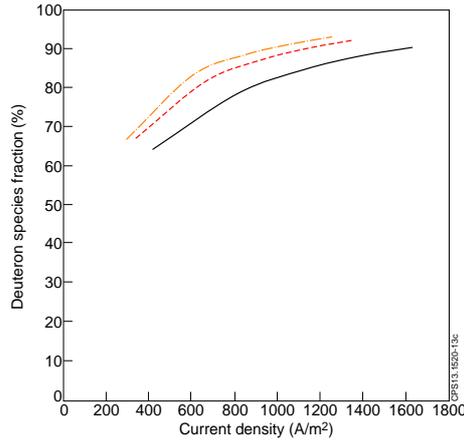

**Figure 6** The effect of increasing the length of the extraction region to increase proton yield by providing a longer path length for molecular break-up. Source lengths are: ▬ 0.2m (standard JET PINI), ▬ ▬ 0.27m and ▬ ▬ ▬ 0.34m. Source pressure is 0.4Pa.

In this figure the arc current was varied up to ~1075A in each case and the only significant difference between the three curves is the lower source efficiency, which moves the curves to the left as the extraction region of the source becomes deeper. This is also shown in figure 7 where the $D_2^+$ and $D_3^+$ fractions are plotted as a function of plasma target, the plasma density integrated along the source length. The three curves in each diagram reflect the three source depths that have been examined but while plasma electrons can break up the $D_2^+$ ions, only the fast primary electrons can destroy the more tightly bound $D_3^+$ ions. The deuteron yield from $D_2^+$ ions benefits slightly from a longer drift region to the filter while the yield from the $D_3^+$ requires a longer (or denser) driver region. However in both cases the extra atomic ion yield is small and is outweighed by the penalties of decreased efficiency and source size.

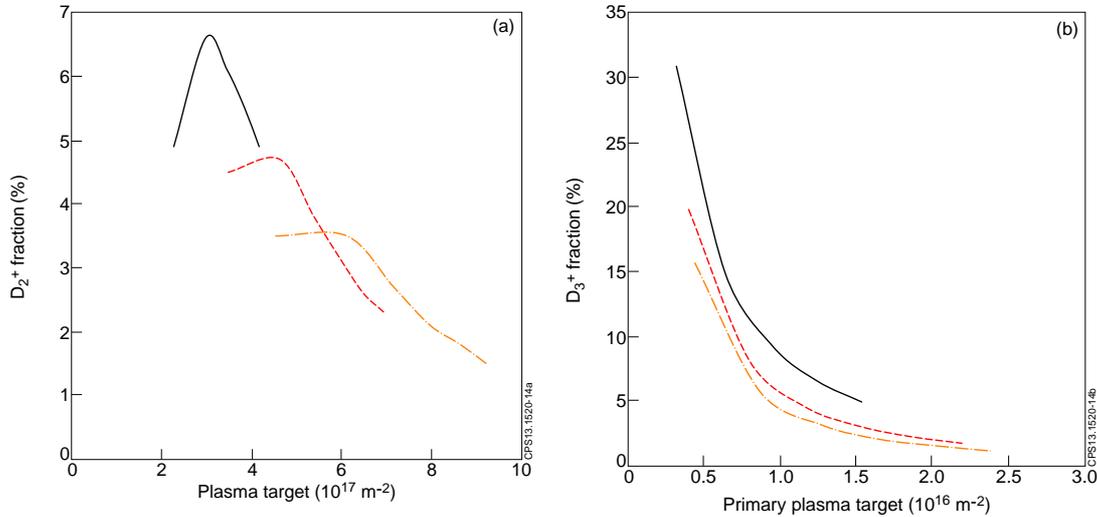

**Figure 7** The influence of plasma target on the molecular ion species for sources with different extraction region lengths. Source lengths are: ▬ 0.2m (standard JET PINI), ▬ ▬ 0.27m and ▬ ▬ ▬ 0.34m. In (a) the $D_2^+$ fraction is shown as a function of the total plasma target. In (b) the $D_3^+$ fraction is shown as a function of the target of primary electrons, their only source of dissociation. Source pressure is 0.4Pa.

*4.2 Ionisation of atomic gas*

In order to isolate the atomic production from the various other pathways a parameter K is defined as:

$$K = \frac{x_0 S_i N_1}{(x_0 S_i N_1 + S_i N_2)}$$

This parameter measures the probability that an ionization collision will lead to a proton being formed as against a $H_2^+$ ion or similarly for a deuterium discharge. $H_3^+$ ions can only be formed from a $H_2^+$ ion collision with a gas molecule. That is the parameter K represents the equivalent of arc current when no molecular ionization occurs. This allows the effect of molecular ionization to be isolated from the analysis.

The code can be manipulated to remove the molecular ion dissociation reactions xs5 and xs7 to eliminate these sources of protons. In the absence of these molecular ion break-up collisions discharges in filter sources and non-filter sources should follow the same trend as the atomic ion can only be formed by a collision between a primary (fast) electron and a gas atom. There are still minor effects such as recombination and changes in the gas temperature (affecting formation of $H_3^+$ and overall source efficiency) but these should be very small. Figures 8 and 9 show the results of the model for such a situation for deuterium and hydrogen discharges in the JET plasma source both with and without a filter.

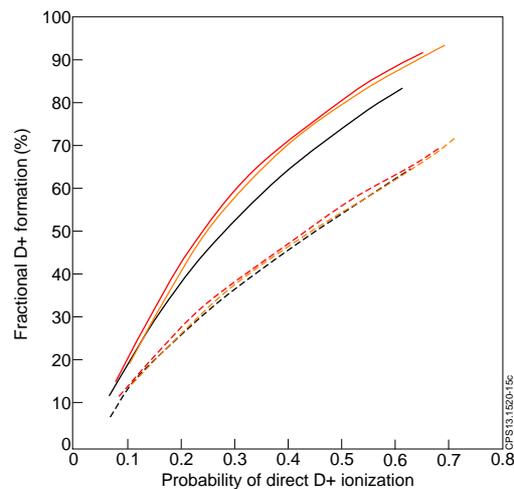

**Figure 8** A deuterium discharge in the JET source with molecular ion dissociation stopped (lower three curves) and the same discharge with dissociation active (upper three curves). The parameter is the magnitude of the filter field, B, as follows: B=0mT with break up ▬, B=1.5mT with break up ▬, B=3.2mT with break up ▬, B=0mT without break up ‒ ‒ , B=1.5mT without break up ‒ ‒ , B=3.2mT without break up ‒ ‒ .

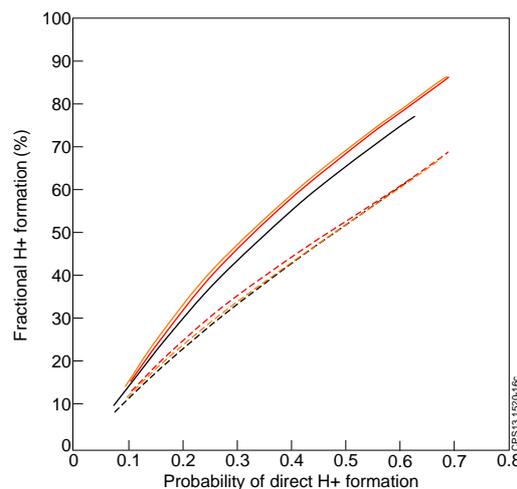

**Figure 9** A deuterium discharge in the JET source with molecular ion dissociation stopped (lower three curves) and the same discharge with dissociation active (upper three curves). The parameter is the magnitude of the filter field, B, as follows: B=0mT with break up ▬, B=1.5mT with break up ▬, B=3.2mT with break up ▬, B=0mT without break up ‒ ‒ , B=1.5mT without break up ‒ ‒ , B=3.2mT without break up ‒ ‒ .

In the absence of molecular ion dissociation, the proton (deuteron) fractions in discharges with 0, 15 and 32 gauss filters are almost the same when plotted as a function of the ionization probability, K. The slight differences arise from gas heating effects and also from the fact that molecular recombination is still active. Roughly two thirds of all protons (deuterons) are formed by direct ionization of the atomic gas at all values of K.

However when the dissociation of molecular ions is active, two major changes occur when the atomic ion fraction is plotted as a function of K. Firstly the proton and deuteron fractions increase significantly, with deuterium showing a larger increase. This is expected as the deuterium molecular ions are exposed to dissociation collisions for a longer time as they drift towards the plasma extraction grid. Secondly, the discharges in a source with a filter field show a larger increase in $H^+$ than those without a filter, again with deuterium discharges showing the larger increase.

To understand the increase in proton yield it is necessary to examine the dependency of the K parameter on the discharge parameters filter field magnetic flux (the field line integrated through the filter length), arc current and source gas pressure. The value of K as a function of filter flux is shown in figure 10 for deuterium discharges in the JET source, this value rather than the peak filter field determining the magnitude of the effect. This shows that the K value increases significantly when there is a filter, by about ~20% for the standard JET source and a larger increase for the negative ion version of the source. The integrated thickness of the magnetic filter is approximately 62mm and the discharge pressure is 0.4Pa of deuterium. The filter flux of the standard JET source is $9.3 \times 10^{-5}$ T.m.

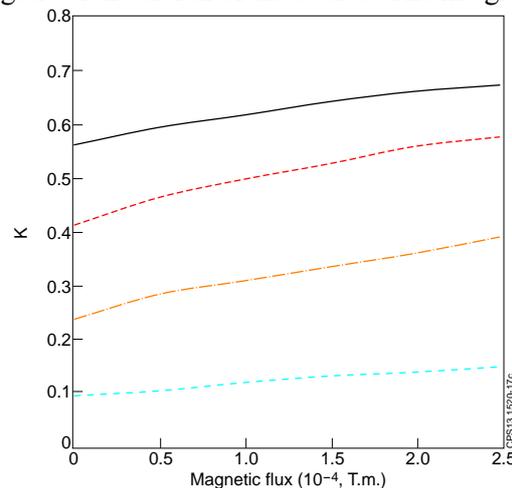

**Figure 10** The dependence of direct ionisation parameter, K, on the integrated magnetic flux along the depth of the source i.e. from driver to extraction regions. The integrated thickness of the magnetic filter is about 62mm and the discharge pressure is 0.4Pa of deuterium. The parameter in arc current: 1000A ▬, 500A ─ ─., 200A ─ • ─., 50A ─ ─. The filter flux of the standard JET source is $9.3 \times 10^{-5}$ T.m

The arc current dependence is shown in figure 11; all processes are active in this plot. This shows the rise of K with arc current for several values of peak filter field, approaching 70% at high arc currents.

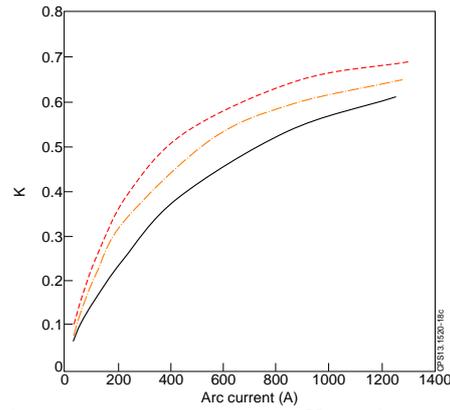

**Figure 11** The dependence of direct ionisation parameter, K, on the arc current. The discharge pressure is 0.4Pa of deuterium. The parameter is the magnitude of the filter field, B: .B=0T ▬, B=1.5mT – • –, B=3.2mT – –

The arc current dependence is shown in figure 11; all processes are active in this plot. This shows the rise of K with arc current for several values of peak filter field, approaching 70% at high arc currents.

Finally the pressure dependence of K is shown in figure 12. This dependence is the weakest in agreement with results from other sources where little dependence of the proton (deuteron) yield on pressure has been seen although the ratio of $D_2^+$ to $D_3^+$ is strongly dependent on pressure. In the case of the JET data the normal operating range is from 0.3Pa to 0.6Pa where K is almost constant.

The filter has a dominant role in determining the value of K itself as seen in figures 2 and 3 for various deuterium discharges in the JET source. Figure 12 shows that the discharge pressure has almost no effect on the value of K as expected unless extremely high or low pressures are considered and even then the effect is weak. Thus the filter acts at one remove on the ion species, not by influencing the break-up of molecular ions as was once thought but instead by affecting the amount of atomic gas in the discharge, mainly via the effects of direct electron dissociation of the gas molecules in the driver region where the density of fast electrons is very high relative to discharges without a filter. This is discussed further in Section 4.4.

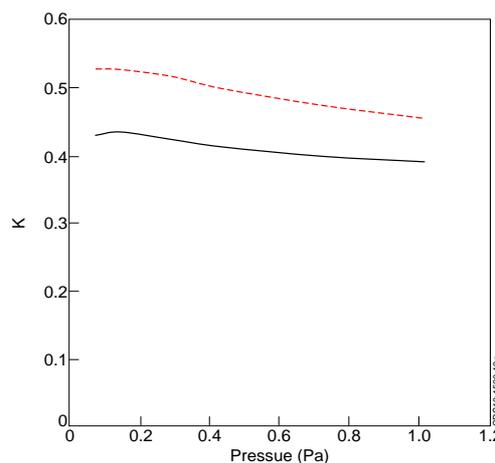

**Figure 12** The dependence of ionisation parameter, K, on the source pressure. The discharge current is 500A in deuterium. The parameter is the magnitude of the filter field, B: .B=0T ▬, B=1.5mT ▬ . ▬

*4.3 Effect of molecular dissociation*
Figures 8 and 9 show that dissociation of the neutral molecule accounts for about 75% of all proton or deuteron production at all values of K but the additional molecular ion dissociation is larger for deuterium discharges. The value of K depends weakly on pressure and filter flux but increases with arc current. However, despite isolating the effects of molecular ionization and dissociation on the species ratios, the discrepancy observed in figures 8 & 9 between the proton yields at different filter field strengths remains to be explained. To investigate further, the data is plotted in a manner that minimises the dominant effect of atomic ionization by maintaining K constant. The combined conversion rate, G, of both molecular ions to atomic ions is given by:

$$G = (n_e \times xs5 + N_2 \times xs6) + (n_f \times xs7 - N_2 \times xs6)$$

As most of the reactions occur in the driver region, the driver values are used in determining the value of G. The ionization of $D_2^+$ is determined by the driver plasma electron density whereas the ionization of the $D_3^+$ ion is determined by the driver primary electron density, so G depends upon the ratio of the two densities. The dependencies of the values of $n_e$ and $n_f$ on the magnetic filter need to be examined. The conversion of $D_2^+$ to $D_3^+$ by collisions with gas molecules cancels and hence the gas density is not involved, which simplifies the analysis.

The only method of modifying the value of G while holding K at a constant value (K = 0.5 is chosen here) is to use the magnetic filter flux as an independent variable. However the filter field alters many of the discharge variables, so the arc current is also used to restore K to constancy at any particular filter flux configuration. The required arc current is shown in figure 13.

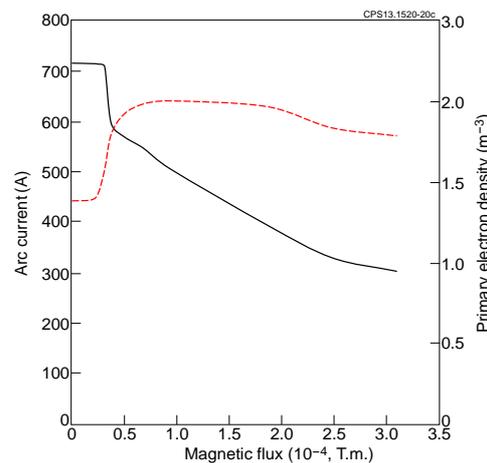

**Figure 13** Plot of the arc current ▬ and the G parameter ▬ ▬ as a function of the integrated magnetic filter flux when K is held constant at 0.5. Deuterium pressure = 0.4Pa.

Note that the flux becomes double valued when plotted against G which arises from the sharp drop in arc current to hold the K parameter constant at increasing filter fluxes. Using the data in figure 13, the plasma electron density, $n_e$, and the primary electron density, $n_f$, in the driver region are plotted as functions of the filter field flux for a deuterium discharge in figure 14. The effect of the magnetization of the electrons is evident as the increase in $n_f$ at the value of $\int Bdl$ = 1.8x10$^{-5}$Tm. Very weak filter fields have almost no effect as the electrons do not become magnetised until the collision frequency equals the cyclotron frequency. This affects the primary electrons first as their collision frequency is much lower than the cooler plasma electrons. At higher filter fields, the reduction in arc current (to keep K constant) becomes large so the primary electron density starts to decrease, despite the trapping action of the filter itself. The plasma electron density simply becomes a constant value.

An ion species model – isotope effects

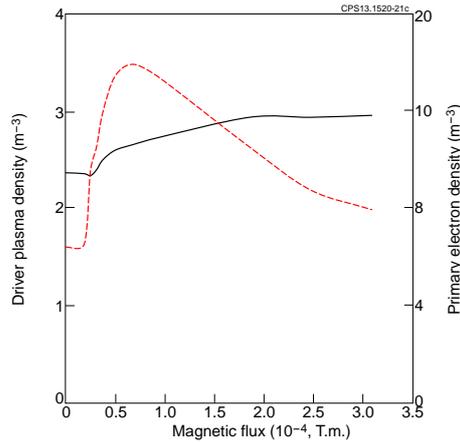

**Figure 14** Plasma electron ▬ and primary electron ▬ ▬ densities in the driver region of a deuterium discharge with a pressure of 0.4 Pa and K=0.5. The influence of the filter field is evidenced by the sharp increase in the primary electron density at an integrated filter filed of $1.8\times10^{-5}$T.m.

Using the magnetic flux as an independent variable, there is a value of G corresponding to each flux value. However G does decrease at high magnetic fluxes as seen in figure 13 and this is due to the decrease in arc current and hence primary density. Plotting the proton (deuteron) fraction as a function of G derived from the variation of driver plasma and primary electron densities with the integrated field, shows a fairly linear tendency as seen in figure 15. The point at a zero value of G is found by switching off molecular dissociation; it still lies on the same trend line as the rest of the data with dissociation active. Note there is no intermediate data for values of G between 0 and $6\times10^{4}$ in hydrogen and 0 and $1.4\times10^{5}$ in deuterium because the molecular dissociation must be either on or off. For deuterium the value of G has a maximum value at about $9\times10^{-5}$ T.m filter flux and once this is exceeded, the deuteron fraction then decreases again with increasing magnetic flux but a falling value of G as seen from figure 12 but still follows the same linear trend with G. The slight difference between the upward and downward arms of the plot arises from the exclusion of the extraction region contribution in the definition of G.

Figure 15 also shows the variation in hydrogen with the parameter, G, which exhibits a similar trend to deuterium. The only major difference is the lower proton yield and the lower general value of G, which arises from the lower plasma densities in a hydrogen discharge and also the fact that the plasma temperatures are not identical. The two data sets lie on a concurrent line due to the identical values of cross sections in xs5 and xs7. The proton and deuteron fractions at equal values of K *and* no molecular ion dissociation are almost identical.

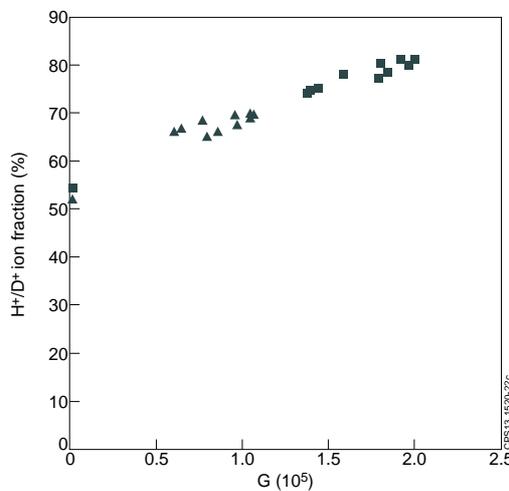

**Figure 15** A plot of deuteron ■ and proton ▲ fractions versus the G parameter characterising molecular dissociation. In this plot the source pressure is 0.4Pa and K = 0.5.

Thus the effect of varying the magnetic flux on molecular ion dissociation is very limited. If the parameter, K, is held constant and likewise the pressure, then using the magnetic filter as a tool to explore the range of molecular ion contribution to the deuteron or proton yield shows that roughly a 10% improvement can be achieved as seen in figure 15, i.e. from~66% to 69% in hydrogen and from 74% to 81% in deuterium. Even artificially de-activating molecular ion dissociation only reduces the deuteron yield by a further 10%. The majority of deuteron production comes from atom ionization after molecular dissociation via the K parameter. However K does show a moderate improvement with the magnetic filter as seen in figures 10 and 11, so the filter does have a purpose here.

**5. Application to tritium injection**
The JET NBI sources make a useful contribution to the diagnostics and fuelling of the JET tokamak plasma during operation in deuterium-tritium mixes. There have been two D-T campaigns on JET in 1997 and 2003 and a third has been proposed in support of ITER to take place around 2016. During the 2003 campaign two PINIs with supercusp magnetic filter field were operated in tritium gas and the beam power measured on the calorimeter [17] was cross calibrated using deuterium data that had been validated by a technique using the stored plasma energy [18]. Since that time, the JET PINIs have been modified to remove the filter field, thus reducing the full energy ion fraction in the beam in order to improve the neutralization efficiency and injected power [19]. As there is no tritium compatible test stand available the species ratio produced by the new, filterless or "chequerboard" source and hence the predicted injected beam power can only be estimated by use of the model. A confident prediction of beam power is essential to planning of plasma experiments when the quantity of tritium is limited.

*5.1 Determination of Accommodation coefficients*
As described in section 3 the accommodation coefficient is a function of many variables and, given its dependency on local surface conditions, is probably unique to an individual source and time. Nevertheless the data analysed in support of the model validation encompasses a number of individual sources of ostensibly the same design, so the values obtained should represent an average. From table 3 and theory [16], a linear extrapolation to tritium implies values of $\gamma=0.2$ and $\gamma_h=0.42$.

*5.2 Predicted values of tritium species ratios*
The results of the modelling for the JET "chequerboard" and "supercusp" sources, operating in tritium, are shown in figures 16 (a) and (b) (species ratios) and figures 17 and 18 (source efficiency). In each case the source neutral pressure is adjusted to accommodate the losses represented by the beam equivalent gas flow, so the pressure varies from 0.7Pa to 0.3Pa as the arc current increases. The maximum triton yield calculated for the supercusp source is 94%. Although there are no measurements of the species ratios in tritium, some degree of confidence can be obtained from the power measurements. The beam power after the neutraliser depends on the efficiency of neutralisation in the gas target and this is a function of neutralizer gas pressure and the beam ion velocity. As the molecular ions in the beam are dissociated into fractional energy atomic ions and hence more easily neutralised, the neutral beam power depends on the initial species ratios. The cross calibration of the tritium neutral bream power with that of deuterium (where the species ratios are known) and application of the neutralization model of Surrey [18] allow an estimate of the species ratios to be made. The result is similar to that from the ion source model – any significant variation would result in a different neutral power at the calorimeter. For the chequerboard source, the computed maximum triton yield is 84%, compared to deuteron yield of 74% in deuterium and proton yield of 60% in hydrogen, presumably due, in part, to the reduce transit speed of the heavier ions leading to increased molecular dissociation.
The source efficiency is almost identical in all three isotopes for both the modelling results and for the measured data, although the model over estimates the extracted current by ~25% at high arc currents in the chequerboard source but under estimates by ~10% for the supercusp [3]. Thus the anticipated extracted current density in the chequerboard source is probably 1800Am$^{-2}$, corresponding to a beam current of 45A in tritium. For the supercusp source the model gives a beam current of 32A at 800A arc current, corresponding to the experimental data of 30A. (The limitation of the achievable arc current was attributed to gas starvation of the source [17]).

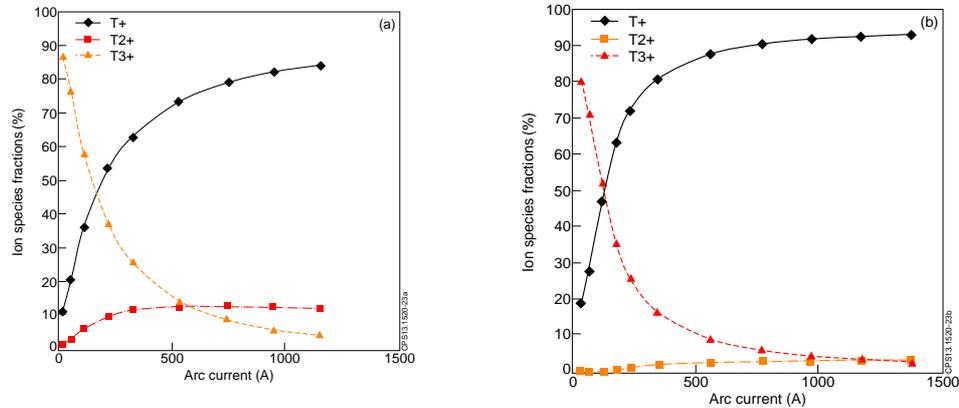

**Figure 16** Species ratios in tritium calculated by the model for (a) the JET chequerboard source and (b) the JET supercusp source. Source neutral pressure varies from 0.7 to 0.3Pa as the extracted beam current increases with arc current

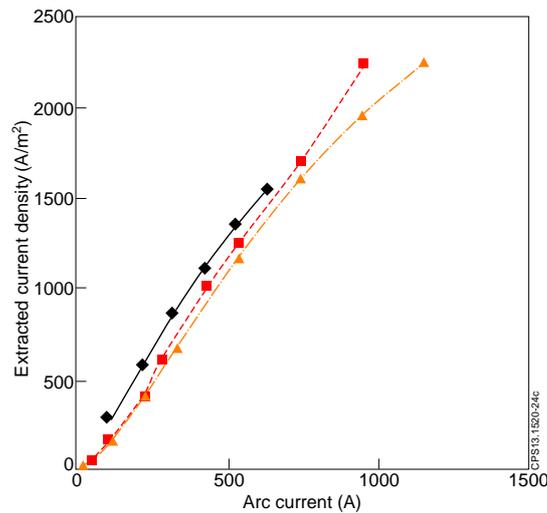

**Figure 17** Source efficiency for the JET chequerboard source calculated by the model operating in hydrogen ♦, deuterium ■ and tritium ▲

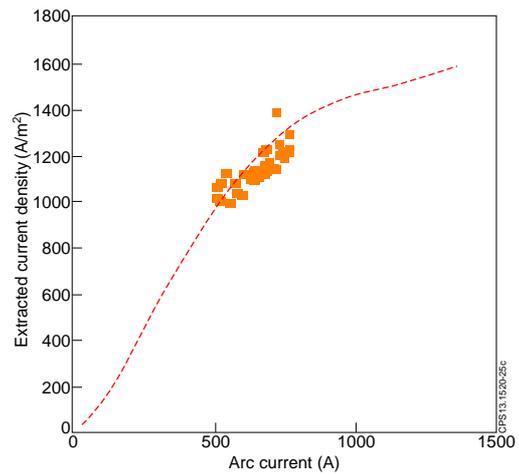

**Figure 18** Source efficiency for the JET supercusp source operating in tritium: data ■ and calculated by the model ▬

Thus the characteristics of the JET chequerboard source can be summarised as shown in table 4

**Table 4** Operating Characteristics of JET Chequerboard Source in Tritium.

| Arc current (A) | Extracted current (A) | T$^+$ fraction | T$_2^+$ fraction | T$_3^+$ fraction |
| --- | --- | --- | --- | --- |
| 800 | 32 | 0.79 | 0.12 | 0.09 |
| 1200 | 45 | 0.84 | 0.12 | 0.04 |

# 6 Conclusions
In conclusion a model of the operation of magnetic multipole sources presented in [3] has been used to investigate the contribution to the proton (deuteron) yield of different dissociation pathways. The model allows certain pathways to be disabled or kept constant, enabling the individual contributions to be assessed. Clearly this would not be possible in experimental devices and illustrates the analytical value of relatively simple models.

This approach has demonstrated that the dissociation of molecular ions is not the major contributor to proton yield but that ionisation of the atomic hydrogen plays a more significant role. If a magnetic filter field is present, conditions in the driver region of the source are modified by magnetisation of the electrons and the primary (fast) electrons in particular increase the molecular dissociation rate and hence produce an increase the proton yield. Increasing the source depth does not lead to a significant increase in proton yield, contrary to earlier assumptions.

Finally the model has been used to predict the species fractions for the JET PINI sources operating in tritium without s magnetic filter field. This shows only a modest reduction in the triton yield compared to that for the supercusp source.

**Acknowledgement**
This work was funded by the RCUK Energy Programme under grant EP/I501045 and the European Communities under the contract of Association between EURATOM and CCFE. The views and opinions expressed herein do not necessarily reflect those of the European Commission. To obtain further information on the data and models underlying this paper please contact PublicationsManager@ccfe.ac.uk